\documentclass[twocolumn,tighten,times]{aastex62}

\usepackage{amssymb,amsmath}
\usepackage[inline]{enumitem} % inline listings 
\usepackage{soul}

\received{}
\revised{}
\accepted{}
\submitjournal{ApJL}

\shorttitle{}
\shortauthors{Syntelis \& Antolin}
\begin{document}

\title{Kelvin-Helmholtz instability and Alfv\'enic vortex shedding in solar eruptions}

\correspondingauthor{P. Syntelis}
\email{ps84@st-andrews.ac.uk}

\author[0000-0002-6377-0243]{P. Syntelis}
\affiliation{St Andrews University,
Mathematics Institute,
St Andrews KY16 9SS,
UK}

\author[0000-0003-1529-4681]{P. Antolin}
%\altaffiliation{Now at: Department of Mathematics \& Information Sciences, Northumbria \\ 
%University, Newcastle Upon Tyne, NE1 8ST, UK}
\affiliation{St Andrews University,
Mathematics Institute,
St Andrews KY16 9SS,
UK}
\affiliation{Now at: Department of Mathematics, Physics and Electrical Engineering, Northumbria University, Newcastle Upon Tyne, NE1 8ST, UK}

\begin{abstract}
We report on a three-dimensional MHD numerical experiment of a small scale coronal mass ejection (CME) -like eruption propagating though a non-magnetized solar atmosphere. 
We find that the Kelvin-Helmholtz instability (KHI) develops at various but specific locations at the boundary layer between the erupting field and the background atmosphere, depending on the relative angle between the velocity and magnetic field. KHI develops at the front and at two of the four sides of the eruption. 
KHI is suppressed at the other two sides of the eruption. 
We also find the development of Alfv\'enic vortex shedding flows at the wake of the developing CME due to the 3D geometry of the field. Forward modelling reveals that the observational detectability of the KHI in solar eruptions is confined to a narrow $\approx10^{\circ}$ range when observing off-limb, and therefore its occurrence could be underestimated due to projection effects.
The new findings can have significant implications for observations, for heating and for particle acceleration by turbulence from flow-driven instabilities associated with solar eruptions of all scales.
\end{abstract}

\keywords{instabilities -- plasmas -- Sun: activity -- Sun: coronal mass ejections (CMEs)--
                Sun: magnetic fields -- magnetohydrodynamics (MHD) -- methods: numerical
               }

\section{Introduction}

KHI can occur at the boundary layer between two moving fluids, when the velocity shear between them is such that it overcomes the surface tension of the boundary layer. 
% In a magnetised environment the KHI has lower growth rates and can be inhibited by magnetic tension when the shear flow has a significant field-aligned component. 
KHI is a very common physical process and has been associated with many astrophysical systems, such as, for example, planetary magnetospheres \citep{Hasegawa_etal2004,Slavin_2010,Masters_etal2010}, astrophysical jets \citep[e.g.][]{Ferrari_etal1981,Bodo_etal1994} and disks \citep[e.g][]{Balbus_etal2003,Johansen_etal2006}. 
In solar related phenomena, KHI has been observed at the flanks of CMEs \citep{Foullon_etal2011,Foullon_etal2013, Mostl_etal2013}, in the dimming regions associated with solar eruptions \citep{Ofman_etal2011}, at the sides of solar jets \citep{Kuridze_etal2016,Li_etal2018, Li_etal2019} and at coronal streamers \citep{Feng_etal2013}.
KHI can play a very important role in the development of turbulence, the mixing and heating of plasma \citep[e.g.][]{Heyvaerts_etal1983,Terradas_etal2008, Antolin_etal2016, Fang_2016ApJ...833...36F} and particle acceleration \citep{2018arXiv181004324L}. 

The threshold and growth rate of KHI at the boundary layer between two horizontal magnetized flows was derived analytically by \citet{Chandrasekhar_etal1961}. The theory has been since extended for flux tubes in the solar convection zone \citep{Tsinganos_1980}, for boundaries between a twisted flux tube and a twisted or a horizontal external field \citep{Zaqarashvili_etal2014a,Zaqarashvili_etal2014b,Zhelyazkov_etal2015a,Zhelyazkov_etal2015b,Zaqarashvili_etal2015}, and also for the effects of partial ionization \citet{Martinez-Gomez_etal2015}. A major point from the theory is that the onset and the growth rate of KHI depends on the angle between the shear flow and the magnetic field at the boundary layer. KHI can develop easier in a shear flow perpendicular to a magnetic field, whereas KHI can be suppressed in a shear flow parallel or antiparallel to a magnetic field \citep[e.g.][]{Hillier_2019MNRAS.482.1143H}.

KHI in the context of solar eruptions has been studied using models that describe shear flows between two vertical regions, mimicking locally the boundary layer of the flank of a CME \citep[][]{Ofman_etal2011, Nykyri_etal2013,Mostl_etal2013,Gomez_etal2016}. In another approach, KHI has been studied by assuming magnetic cylinders moving through a model atmosphere, mimicking the ejection of a flux rope \citep{Pagano_etal2007}. Such models have the advantage of being easy to implement. As such, they are useful to perform controlled parametric studies. However, the magnetic field of a full 3D CME eruption is significantly more complicated than such idealized configurations \citep[e.g.][]{Syntelis_etal2017}.  
In this study, we report on the first 3D MHD simulation of a solar eruption showing the development of KHI at the boundary layer between the erupting field and the background atmosphere. 
We show that the KHI only develops at specific boundaries where the shear flow is mostly perpendicular to the local magnetic field, with important implications for assessing its occurrence as well as the generation of turbulence during CME propagation.

\begin{figure*}
    \centering
    \includegraphics[width=\textwidth]{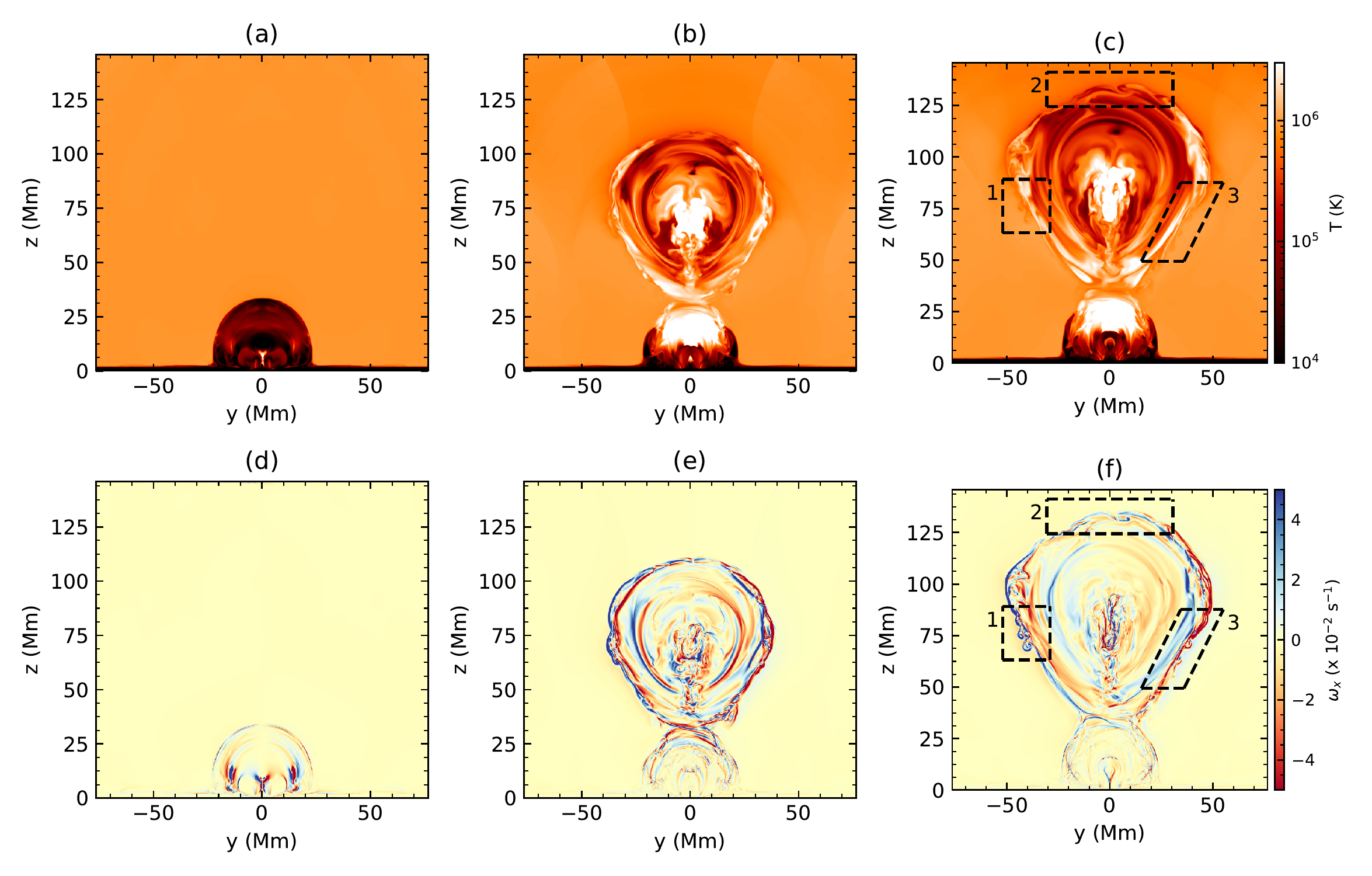}
    \caption{ 
        Time time evolution of temperature (top row) and $x$-component of the vorticity (bottom row) at the $yz$-midplane. 
        The first, second and third columns correspond to $t=107.8, \,128.6, \,135.7$~min.
        The boxed regions in (c), (f) show regions where KHI has developed. 
    }
    \label{fig:time_evol}
\end{figure*}
\begin{figure*}
    \centering
    \includegraphics[width=\textwidth]{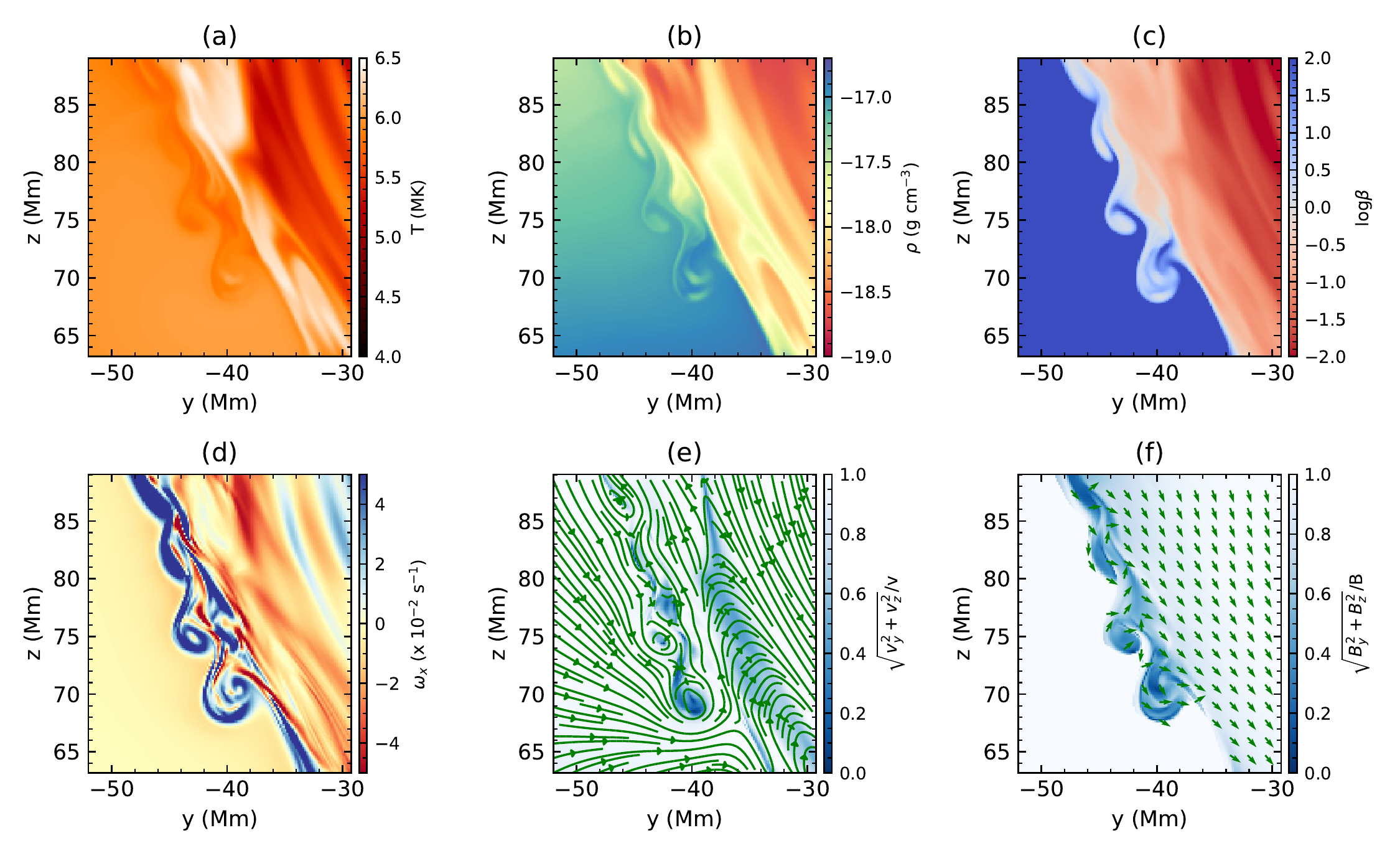}
    \caption{ 
        A close-up of region 1, shown in Fig.~\ref{fig:time_evol}. Panel (a) shows temperature, (b) density,  (c) plasma $\beta$, and (d) $x$-component of the vorticity.
        (e) The magnitude of the planar velocity over the total velocity. Over-plotted are streamlines of the planar velocity field. 
        (f) The magnitude of the planar magnetic field over the total magnetic field. Over-plotted is the planar magnetic field vector.
    }
    \label{fig:region1}
\end{figure*}

\section{Model}
\label{sec:initial_conditions}

The model used in this paper is described in detail in \citet{Syntelis_etal2017,Syntelis_etal2019}. 
We use Lare3d \citep{Arber_etal2001} to solve the 3D time-dependent, resistive and compressible MHD equations in Cartesian geometry.
The equations, resistivity form and normalization can be found in \citet{Syntelis_etal2017}. 

The numerical domain has a physical size of 153$^3$~Mm (1000$^3$ grid points). 
The boundary conditions are periodic in the $y$-direction and open in the $x$-direction. In the $z$-direction the boundaries are open (closed) for the top (bottom) of the domain.

The domain consists of a solar interior and a solar atmosphere. The interior is a convective stable and adiabatically stratified layer between $-7.2\ \mathrm{Mm}\le z < 0 \ \mathrm{Mm}$. 
The atmospheric temperature follows a hyperbolic tangent profile for the temperature, mimicking an isothermal photospheric-chromospheric layer ($0 \ \mathrm{Mm} \le z < 1.8 \ \mathrm{Mm} $), a transition region ($1.8 \ \mathrm{Mm} \le z < 3.2 \ \mathrm{Mm}$) and an isothermal corona ($3.2 \ \mathrm{Mm} \le z < 145.8 \ \ \mathrm{Mm}$).
The atmospheric density and pressure are derived by assuming hydrostatic equilibrium.

A horizontal flux tube is positioned inside the solar interior, along the $y$-direction, at $z=-2.1 \ \mathrm{Mm}$.
The flux tube's magnetic field in cylindrical coordinates is:

\begin{align}
B_{y} &=B_\mathrm{0} \exp(-r^2/R^2), \\
B_{\phi} &= \alpha r B_{y}
\end{align}
where $R=450$~km is the tube's radius, $r$ the radial distance from the tube axis, $\alpha=0.0023$~km$^{-1}$ is the twist tube and  $B_0$=2400~G is the magnetic field strength at its center.
The flux tube is initially in pressure equilibrium. To initiate the emergence of the flux tube a density deficit is imposed along the tube's axis \citet[e.g.][]{Fan_2001}:

\begin{equation}
\Delta \rho = \frac{p_\mathrm{t}(r)}{p(z)} \rho(z) \exp(-y^2/\lambda^2),
\label{eq:deficit}
\end{equation}
where $p$ is the external pressure,  $p_\mathrm{t}$ is the total pressure within the flux tube and $\lambda=0.9$~Mm is the length scale of the buoyant part of the flux tube.

The formation of KHI depends on the relative angle of the shear flow and the direction of the magnetic field. 
We assume a non-magnetized background atmosphere to assess the locations at the surface of the eruption where KHI can develop, solely due to the direction of the erupting field, without the influence of a preferred direction imposed by an ambient field. 
The inclusion of an ambient field can potentially affect the locations where KHI develops and make KHI less prominent \citep[e.g.][]{Pagano_etal2007} (see Discussion).

The real viscosity in the solar corona is estimated to be many orders of magnitude smaller than the numerical viscosity in our simulation. Unrealistically high viscosities can suppress the formation of KHI by significantly reducing the velocity shear between the two interfaces \citep[e.g.][]{Howson_etal2017b,Antolin_etal2019}.
It is possible that with higher spatial resolution and correspondingly lower numerical viscosity, additional KHI could be obtained in our model.
In this study, we have not examined the effects of resolution and numerical viscosity to the development of KHI as it would be very demanding computationally.

\section{Results}
\label{sec:results}

Figure~\ref{fig:time_evol} shows the time evolution of the temperature (first row) and $x$-component of the vorticity (second row) before and during the eruption. 
The top-left panel shows the temperature distribution at the vertical midplane, inside which a low-lying flux rope is located. This flux rope has been formed by the low-lying gradual tether-cutting reconnection of sheared lines \citep[details in ][]{Syntelis_etal2017,Syntelis_etal2019}.
During the eruption (middle column), the erupting field (structure between, $30<z<110$~Mm, (b)) develops a hot edge layer (outer part of the field).
We will refer to the outer-most part of the edge as the boundary layer between the  magnetized eruption and the non-magnetized atmosphere.
The temperature at the very edge of the boundary layer is cooler than that of the background atmosphere, and cooler than that of the hot edge part of the eruption. This is due to the adiabatic expansion of this part of the boundary layer, as the field strength there decreases to zero.

During the eruption, the erupting field displaces the atmospheric material above it. The displaced material is pushed away from the upper-most part of the erupting field and then down besides the sides of the eruption.
Therefore, velocity shear develops at the boundary layer, both by the upwards velocity of the erupting field and by the sideways and downwards velocity of the displaced material. This velocity shear can be seen as regions of high vorticity (Figure~\ref{fig:time_evol}(d)-(f)).

KHI develops at three locations at the boundary layer ((c), (f)). Regions 1, 3 are the left and right flank of the eruption when observing the structure along the $+x$-direction. Region 2 is the upper part of the erupting field. There, KHI vortices are bigger in size.

\begin{figure*}
    \centering
    \includegraphics[width=0.9\textwidth]{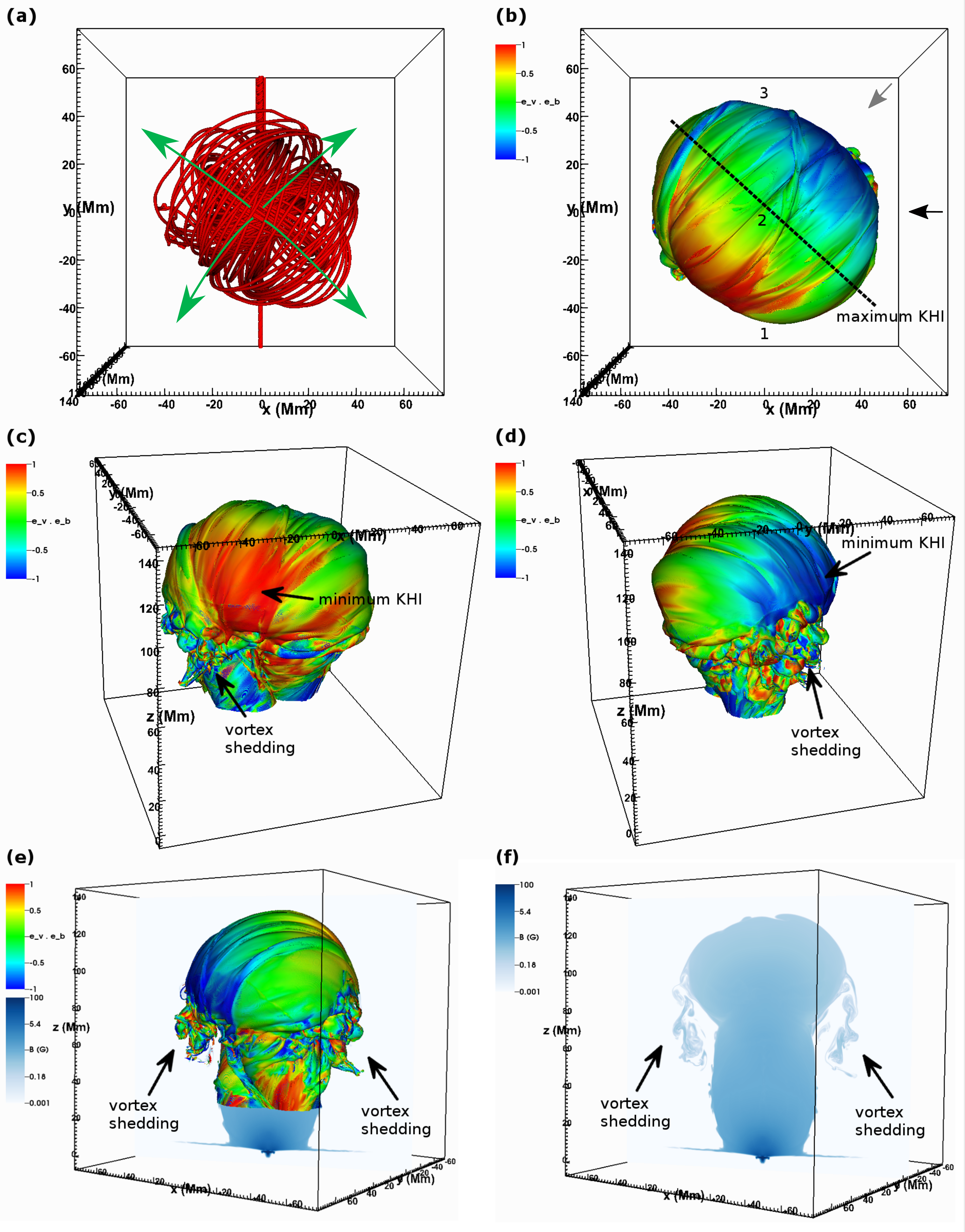}
    \caption{ 
        (a) Sample field lines of the eruption (red). Green arrows indicate the  direction of the velocity field in the $xy-$plane.
        (b) Isosurface of vorticity at $|\omega|=1.75 \times 10^{-2} s^{-1}$, colored by $\mathbf{\hat{v}}\cdot\mathbf{\hat{B}}$. Dashed line shows the region where $\mathbf{\hat{v}}\cdot\mathbf{\hat{B}}\approx0$ (maximum KHI). Labels 1, 2, 3 indicate the approximate location of regions 1, 2, 3 of Figure~\ref{fig:time_evol}(f).
        Black and grey arrows indicate the line-of-sights used in Figure~\ref{fig:fomo}.
        Panels (c), (d) are side views of panel (b). Arrows indicate regions where $\mathbf{\hat{v}}\cdot\mathbf{\hat{B}}\approx\pm1$ (minimum KHI) and regions of vortices associated with vortex shedding.
        (e) Side view of Figure~\ref{fig:3dkhi}(b), better for visualizing the vortex shedding regions.
        (f) Magnetic field strength at a vertical slice through the vortex shedding regions (also shown in (e)).
    }
    \label{fig:3dkhi}
\end{figure*}

Region 1 is examined in more detail in Figure \ref{fig:region1}. 
The temperature (a) and density (b) of the wave-like features take values between that of the background atmosphere and that of the boundary layer prior to the development of the waves (Figure~\ref{fig:time_evol}b). However, KHI has not yet developed enough to mix the hot edge of the eruption (right of the wave-like features) with the atmosphere (left of the wave-like features).
This is similar to the initial stage of the mixing found in simulations of KHI in coronal loops due to transverse MHD waves \citep[e.g.][]{Antolin_2014ApJ...787L..22A, Howson_etal2017, Howson_etal2017b,Karampelas_etal2017}.
The plasma $\beta$ inside the vortices, although mostly high (blue/white, panel (c)), increases locally at the fronts of the waves (white/red), indicating the local compression of the plasma due to the instability.

Panel (d) shows the $\omega_x$ component of the vorticity. The wave-like features have mostly positive $\omega_x$ (blue), indicating a consistent counter-clockwise rotations in the $yz-$plane. This rotation is also reflected in the streamlines of the planar velocity field (green lines, (e)). 
The negative values of $\omega_x$ (red) near the base of the wave-like features are associated mostly with shearing and not with an actual rotation.

\begin{figure*}
    \centering
    \includegraphics[width=\textwidth]{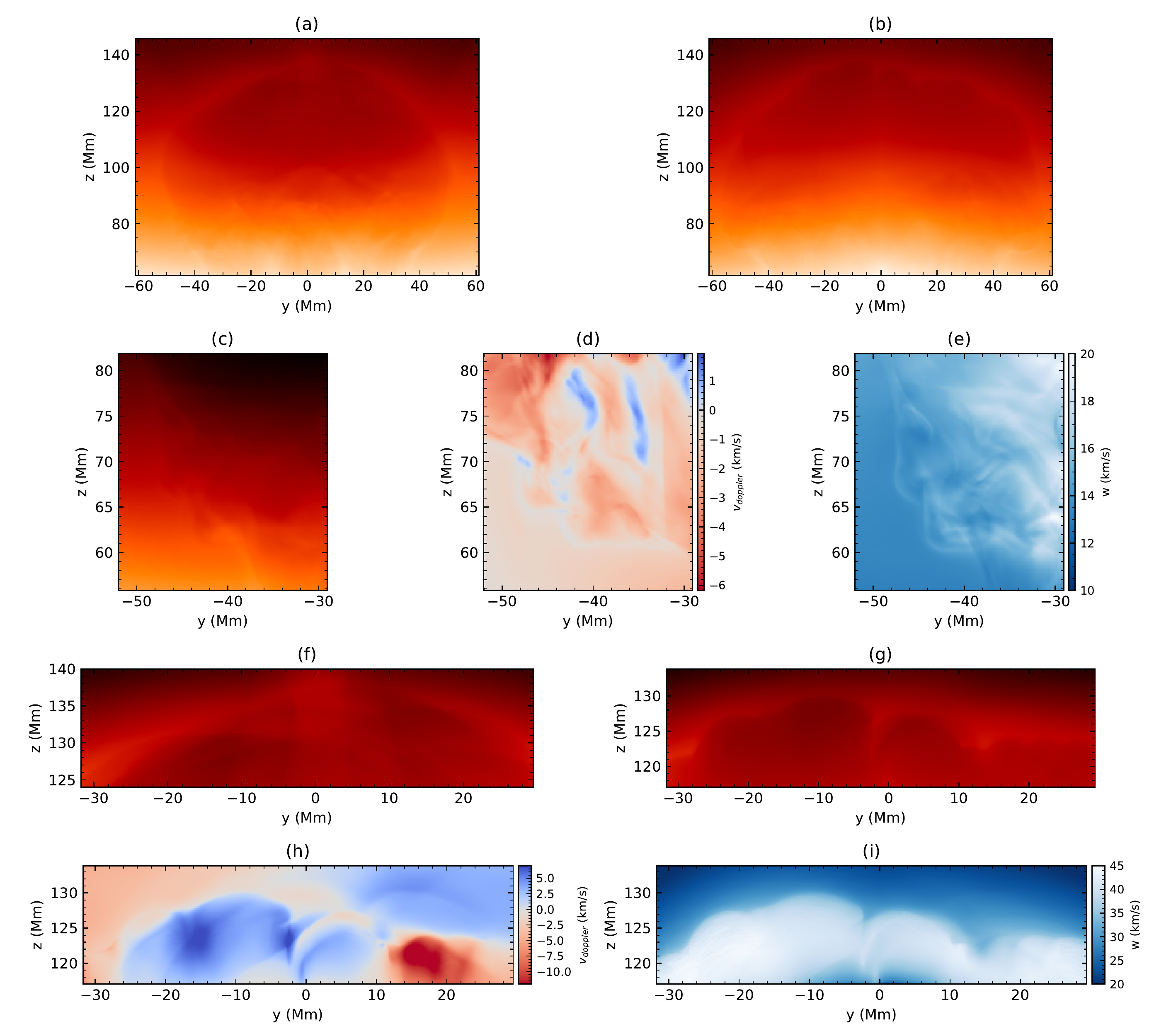}
    \caption{
        Synthetic observations of \ion{Fe}{9} for two line-of-sights, one  perpendicular to the plane of Figure~\ref{fig:time_evol} ($x$-axis, 90\degr~polar angle, panels (a), (c), (d), (e), (f)) and another approximately perpendicular to the maximum KHI line of Figure~\ref{fig:3dkhi}(b) (45\degr~polar angle, panels (b), (g), (h), (i)).
        (a), (b) Line intensity showing the whole eruption. 
        (c), (d), (e) Line intensity, Doppler velocity and line width of region 1.
        (f) Line intensity of region 2 at 90\degr~angle.
        (g, h, i) Line intensity, Doppler velocity and line width of region 2 at 45\degr~angle.
    }
    \label{fig:fomo}
\end{figure*}

Inside the central part of two lower wave-like features ($z\approx70-75$~Mm), $\omega_x$ changes sign. This is due to the off-plane motion of the vortices, indicating that the KHI vortices have a 3D structure.
To examine that off-plane motion, we plot the ratio of the magnitude of the planar velocity over the magnitude of the total velocity (panel (e)). White indicates a planar flow and dark blue indicates a flow away from the plane. 
Similarly, in (f) we plot the planar magnetic field strength over the total magnetic field strength.
The velocity field shows a component away from the plane, contrary to the magnetic field, implying that the velocity shear is at an angle locally to the magnetic field, therefore favoring the development of KHI.
Also, the KHI features develop internally a 3D structure where the velocity, magnetic and vorticity fields have strong off-plane components. 
We now examine the 3D structure of the whole boundary layer of the eruption (Figure~\ref{fig:3dkhi}). In (a), we trace magnetic field lines from the edges of the eruption (not the KHI regions). The eruption displaces the material above it, and moving it away from the apex of the eruption, towards every direction (green arrows). In some locations the flow is perpendicular to the magnetic field (favouring KHI), while in other locations the flow is parallel or anti-parallel to the field (not favouring KHI).
To visualize this, in (b) we plot the isosurface of the magnitude of vorticity colored by $\mathbf{\hat{v}}\cdot\mathbf{\hat{B}}$, where $\mathbf{\hat{v}}$ and $\mathbf{\hat{B}}$ are the unit vectors in the direction of the velocity and the magnetic field. 
Green indicates the locations favouring KHI, while red and blue indicate the locations not favouring KHI. 
The 3D visualization reveals that a significant part of the surface of the eruption is covered with wave-like features of different sizes, thereby corrugating the surface. These features have maximum amplitude approximately along the dashed line, where $\mathbf{\hat{v}}\cdot\mathbf{\hat{B}}\approx0$. The numbers 1, 2, 3 denote approximately the locations of the 2D boxes of Figure~\ref{fig:time_evol}.
Moving from the green regions towards the red (c) or blue (d) regions, the amplitudes of the wave-like features decrease and eventually become zero (arrows showing minimum KHI). 
As a result, at the two sides of the eruption associated with the red and blue regions, there are no signs of KHI. 
At the two other sides, the field geometry is such that the KHI develops. 
The complex 3D structure of KHI at various locations on the surface of the eruption implies that projection effects could be highly influencing the observational detection of these features.

\begin{figure*}
    \centering
    \includegraphics[width=\textwidth]{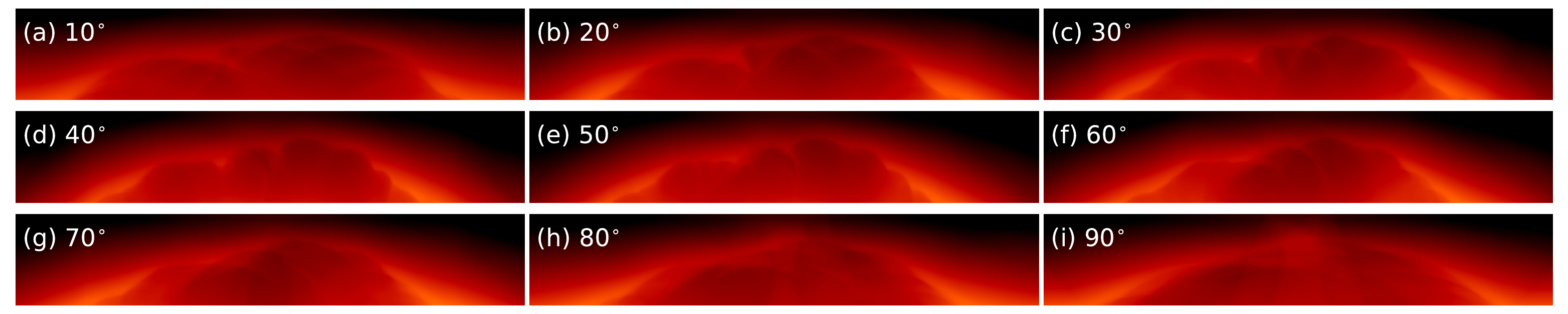}
    \caption{Line intensity of region 2 for line-of-sights between 10\degr and 90\degr.
    }
    \label{fig:projection_effects}
\end{figure*}

To assess the projection effects, we produce synthetic observations of \ion{Fe}{9} along two lines-of-sights using the FoMo code \citep{VanDoorsselaere_etal2016} 
(Figure~\ref{fig:fomo}), one being perpendicular to the plane of Figure~\ref{fig:time_evol} ($x$-axis, 90\degr~angle, see black arrow in Figure~\ref{fig:3dkhi}b).
At that angle regions 2 and 3 show no sign of KHI (a). Panels ((c)-(e)) show the line intensity, Doppler velocity and line width of region 1 (comparable to Figure~\ref{fig:region1}). For simplicity, the spectral features have been obtained by fitting single Gaussians to the spectral profiles. However, the KHI introduces multiple spectral components and the Doppler velocities shown here do not represent the extrema. These are represented by the broad line widths obtained from the single Gaussian fits, which are mostly of non-thermal origin. The KHI features are visible in the intensity image, however, they are more prominent in the Doppler velocity and line width images. 
The second line-of-sight is approximately perpendicular to the maximum KHI line of Figure~\ref{fig:3dkhi}(b) (45\degr~angle, grey arrow). 
Region 2 now shows some undulating features, whereas regions 1 and 3 do not show signs of KHI. Panels (f) and (g) show a close-up of region 2 for the two angles. In (f) there is no indication of KHI. In (g), the undulating feature is present, however, it does not appear as KHI waves, as they are ``smoothed'' due to the integrating effect of the optically thin radiation across the complex region of the top of the eruption. Therefore, observations of such features are likely not to be associated with KHI.
The Doppler velocity (h) and line width (i) at the 45\degr~angle again produce clear signatures of these undulating features, however, without showing characteristic KHI crests. The complexity of region 2 can be witnessed in the line width, where the emission from plasma flowing at different directions results in very wide line profiles with high non-thermal broadening. 

To further assess the projection effects, in Figure~\ref{fig:projection_effects} we show the line intensity of region 2 for nine angles between 10\degr and 90\degr. The undulating features appear more clearly between 40\degr-50\degr and become smoothed away from these angles.
In general, for lower resolution and noisier data, such features would be further degraded. Similar results are obtained for regions 1 and 3.
Therefore, current observations might be underestimating the presence of KHI in solar eruptions due to line-of-sight projection and resolution effects.

Besides features associated with KHI, in Figure~\ref{fig:3dkhi}(c), (d), we find other localized regions of vorticity. These correspond to Alfv\'enic vortex shedding developing at two of the sides of the eruption (``vortex shedding'' arrows).  
These regions are more clearly shown in Figure~\ref{fig:3dkhi}(e). In (f), we plot the magnetic field strength at a vertical slice through the vortex shedding regions of (e). 
At the two sides of (e, f), the field geometry is such that the ``legs'' and the upper part of the eruption form two convex regions. The flow moving around the sides of the upper part of the eruption and towards the narrow point between the convex regions becomes turbulent. The drag forces deform the boundary layer, forming these characteristic thin and turbulent structures associated with vortex shedding \citep{Gruszecki_2010PhRvL.105e5004G, Bonet_2008ApJ...687L.131B, Nakariakov_etal_2009AA...502..661N}.

\section{Discussion}
\label{sec:conclusions}

We have presented results on the formation of KHI and vortex shedding in a simulation of a small scale solar eruption. The results demonstrate that using a realistic 3D model of a solar eruption is crucial to assess the formation and the effects of KHI and other flow-driven instabilities in the local development of turbulence, the heating of plasma, and any possible effects associated with such processes \citep[e.g. small scale reconnection, local acceleration of particles][]{Nykyri_Otto_2004AnGeo..22..935N,Fang_2016ApJ...833...36F,2018arXiv181004324L,Paez_etal2019}.

The reported simulations have been performed in a non-magnetized atmosphere. Adding an ambient field would affect the formation of KHI \citep[e.g.][]{Pagano_etal2007}. Changes on the local plasma properties at the boundary layer (e.g. shear velocity, boundary layer width, relative magnetic field orientation, magnetic field strength, compressibility of the fluid etc), can lead to e.g. changes in the growth rate of the instability and affect the sizes, shapes and mixing of the vortices. Such effects can make the instability less prominent or even suppress it, depending on the specific local properties \citep[e.g.][]{Miura_etal1982,Ryu_etal2000, Nykyri_etal2013,Mostl_etal2013, Tian_etal2016,Faganello_etal2017, Ma_etal2017}.
Also, an ambient field would affect the motions of the displaced material, changing the angle of the shear and thus affecting the locations where KHI can develop.
\citet{Mostl_etal2013} showed that depending on the relative angles between the magnetic field and the velocity shear along both sizes of a horizontal magnetized layer, the layer could become unstable at only one of its two sides. Given the complexity of the 3D magnetic field of a solar eruption (Figure~\ref{fig:3dkhi}(a)), the presence of an ambient field could potentially suppress KHI in one of the two unstable sides of the eruption, allowing KHI to develop only in one side of the eruption, similar to the observations of \citet{Foullon_etal2011}. However, our results strongly suggest that such a case can be easily attributed to projection effects resulting from the line-of-sight of the observation. Such possibilities can only be examined by properly modeling the 3D field of the solar eruption.

On the other hand, we expect Alfv\'enic vortex shedding to be largely independent of the external magnetic field, since its occurrence is solely due to the presence of a sharp obstacle with respect to the flow direction (convex regions in our simulation).

Forward modeling of the eruption revealed that the projection effects associated with optically-thin observations of solar eruptions can significantly underestimate the presence of KHI in solar eruptions. This provides a simple explanation to the scarcity of the detection of KHI features in observations of solar eruptions. On the other hand, corrugated CME fronts produced by KHI as in our model may have easily been disregarded, interpreted simply by differential CME expansion or other effects. Spectral diagnostics could potentially provide better evidence of KHI related features than imaging diagnostics.

The studied eruption has the energy and physical scale of a small scale eruption. Its magnetic field strength is expected to be weaker than that of larger eruption. Therefore, it is possible that KHI develops easier at the periphery of smaller and less energetic events \citep[including jets, e.g.][]{Li_etal2018,Li_etal2019}, at least while the erupting fields are still located close to the solar surface so that they are observed by EUV imagers and spectrometers.
The formation of the KHI in small scale eruptions can be a very useful local and global diagnostic to probe the magnetic field and local plasma properties of these events. This is further so, given that the KHI is highly sensitive to the angle between the velocity shear and the local magnetic field, thereby providing information of the global magnetic field topology.

Similar eruptions to the studied one can be produced in higher energies in a multi-scale manner following a power law distribution \citep{Syntelis_etal2019}.
For larger scale eruptions, the magnetic field could be strong enough so that KHI develops less frequently closer to the solar surface. However, depending on the properties of the eruption, KHI could also develop away from the lower solar atmosphere, during the eruption's expansion and propagation inside the solar wind \citep{Paez_etal2017}, as the shearing between a CME and the solar wind could be appropriate for the formation of KHI \citep{Manchester_etal2005}. Therefore our results should be applicable to larger scale eruptions showing KHI either closer or further away from the solar surface.

The presence of turbulence in the solar wind and the existence of a shock and an expanding CME front is often invoked to explain the acceleration of particles associated with these events \citep{Bemporad_2011ApJ...739L..64B}. The generated turbulence associated with the KHI and Alfv\'enic vortex shedding as well as the resulting corrugated CME front demonstrated in our model should therefore be considered for turbulence and particle acceleration studies.

\acknowledgments

This work was supported by computational time granted from the Greek Research \& Technology Network (GRNET) in the National HPC facility - ARIS. P.A. acknowledges funding from his STFC Ernest Rutherford Fellowship (No. ST/R004285/1). P.S. acknowledge support by the ERC synergy grant ``The Whole Sun''.

\bibliographystyle{aasjournal}
\bibliography{bibliography}

%\clearpage

\end{document}